\documentstyle[11pt,aaspp4]{article}

\begin{document}
\title{Galaxies with Spiral Structure up to $z \approx 0.87$ \\
Limits on $M/L$ and the Stellar Velocity Dispersion 
}

\author{
A.~C.\ Quillen\altaffilmark{1}$^,$\altaffilmark{2}, \&
V.~L.\ Sarajedini\altaffilmark{1}$^,$\altaffilmark{3}
}
\altaffiltext{1}{The University of Arizona, Steward Observatory, Tucson, AZ 85721}
\altaffiltext{2}{E-mail: aquillen@as.arizona.edu}
\altaffiltext{3}{E-mail: vicki@as.arizona.edu}

\def\spose#1{\hbox to 0pt{#1\hss}}
\def\lta{\mathrel{\spose{\lower 3pt\hbox{$\mathchar"218$}}
     \raise 2.0pt\hbox{$\mathchar"13C$}}}
\def\gta{\mathrel{\spose{\lower 3pt\hbox{$\mathchar"218$}}
     \raise 2.0pt\hbox{$\mathchar"13E$}}}

\begin{abstract}
We consider seven distant galaxies with clearly evident spiral structure
from HST images.  Three of these were chosen from \cite{vog96} (VFP)
and have measured rotational velocities.  
Five were chosen from the Medium Deep Survey 
and are studied in \cite{vic} (SGGR), and one galaxy is found in both papers.
We place upper limits on their mass-to-light ratios ($M/L$) by
computing $M/L_B$ for a maximal disk.  
We find that these galaxies have maximal disk mass-to-light ratios 
$M/L_B = 1.5 - 3.5 ~ M_\odot/L_{B\odot}$ at the low end, but within the range
seen in nearby galaxies. 
The mass-to-light ratios are low
enough to suggest that the galaxies contain a young,
rapidly formed stellar population.  

By using a Toomre stability criterion for formation of 
spiral structure, we place constraints
on the ratio of $M/L$ to the stellar velocity dispersion.
If these galaxies have maximal disks
they would have to be nearly unstable so as to have 
small enough velocity dispersions that 
their disks are not unrealistically thick. 
This suggests that either a fraction of the light originates from
a component of stars in a thick disk or bulge not strongly affected by
the spiral structure, 
or there is a substantial amount of dark matter present in the luminous
regions of the galaxy.
\end{abstract}


\section {Introduction}
The Hubble Space Telescope (HST) has observed distant galaxies
with sufficient angular resolution to reveal morphologies
of galaxies on kpc scales. 
Some galaxies at moderate redshift (up to $z \sim 1$)
show spiral structure similar to nearby galaxies.  
Recent spectroscopic studies such as \cite{rix97}, \cite{vog96} and \cite{vog97}
have measured disk rotational velocities for these galaxies, allowing
initial comparisons of the velocity-luminosity relations
(Tully-Fisher) between distant and local galaxies.
In this paper we use the existence of spiral structure evident from 
the HST images coupled with values estimated for their disk rotational
velocities, to constrain the properties of their disks. 
Optical or UV rest frame images are ideally suited for 
study of spiral structure at high redshift because spiral structure
is expected to be more prominent in the bluer bands.

If we assume that the spiral structure that we observe is
due to spiral density waves, then the disk must be
sufficiently responsive or unstable to form these waves.
In other words the Toomre $Q$ parameter (see \cite{B+T}) 
which depends both on the disk mass-to-light ratio ($M/L$)
and the velocity dispersion of the disk,
is expected to be confined to the fairly narrow range of $1< Q < 2$. 
A disk with $Q<1$ is violently unstable and a disk with $Q>2$ has
such a high velocity dispersion
that spiral density waves should not be observed.
Stellar velocity dispersion measurements of nearby galaxies 
are consistent with these stellar disks being close to this instability range 
(\cite{bot93}).
With the above inequality for $Q$, we place limits on the ratio of 
$M/L$ and the stellar velocity dispersion  for the distant galaxies
where HST reveals spiral structure.  We compare 
this product to the mass-to-light ratio required for a ``maximal disk.''  
This ratio determines the most massive disk that could be present 
within the constraint set by the rotation curve.  
We assume that the disk in these galaxies has 
a mass-to-light ratio that does not vary with radius.

\section {Galaxies Chosen}

We consider five galaxies from \cite{vic} (hereafter SGGR; 
also studied in \cite{sar96}), 
observed in the Medium Deep Survey (\cite{gri94}, \cite{rat97}).  
These galaxies were chosen because of their clearly evident, prominent, 
nearly bisymmetric spiral structure seen in the 
HST Wide Field Planetary Camera images observed in the F814W ($I_{814}$) filter 
displayed in Fig.\ 1.  For these galaxies 
the bulk of the light is from an exponential disk with 
contributions of less than a few percent from a bright unresolved nuclear
component (SGGR).

We also consider three galaxies displayed and studied in 
\cite{vog96}, (hereafter VFP),
observed as part of the Deep Extragalactic Evolutionary
Probe (DEEP) project (\cite{koo95}).
These galaxies were also chosen because of their clearly evident, prominent,
nearly bisymmetric spiral structure seen in the HST images.  
Their rotational velocities were measured spectroscopically 
by these same authors.  
We did not choose galaxies from the sub-L$_*$ sample of \cite{vog97} because
none of these galaxies had obvious spiral structure.
All galaxies chosen are brighter than $L_*$ ($M_B \sim -20.4$).
One galaxy (denoted as in 0305-00115 is the same as uem00-4 in SGGR) 
is found in both samples.
We list this galaxy twice in all our tables so as to maintain 
consistency between the measured quantities.  

Observed galaxy and disk parameters for these galaxies are listed in Table 1.
The surface brightness profile assumed by VFP and SGGR 
was that of an exponential disk, with 
exponential scale length denoted here by $h$.
Derived rest frame properties are listed in Table 2.
Rest wavelength $B$ band central surface brightnesses, $S_B(r=0)$,
and absolute magnitudes 
were computed using $H_0 = 75$km s$^{-1}$ Mpc$^{-1}$, 
$q_0=0.50$ and $0.05$, a cosmological constant of zero 
and an internal extinction correction 
based on the galaxy inclination from \cite{rub85}.    
Rest frame magnitudes were computed with a code provided by
C.~Gronwall which fits redshifted model galaxy 
spectral energy distributions to the observed $V-I$ color.
By computing rest frame magnitudes in $B$ band from the observed
$V$ and $I$ band magnitudes we have attempted to minimize errors
in the K-correction for the highly redshifted galaxies.
The central surface brightness is computed from the galaxy luminosity
using $S(r=0) = L / 2 \pi h^2$; this is accurate 
when the disk component dominates.

Values for the circular velocity, $v_{TF}$, predicted
from the Tully-Fisher relation, are listed in Table 3.  We computed
$v_{TF}$ from the $B$ band rest frame absolute magnitudes with the local 
($z=0$) $B$ band Tully-Fisher relation of \cite{pie92}.  
As found by \cite{vog97} the differences between the circular velocities 
predicted
from the local Tully-Fisher relation and those observed for high redshift
galaxies are not extremely large  ($\lta 1$ mag)
though these three galaxies have lower magnitudes than expected from the 
local Tully-Fisher relation.

The physical scale lengths of the galaxies are similar to that 
of the Milky Way ($h \sim 3$ kpc; see summary in \cite{sac97}).
Their rotational velocities (observed or predicted by the Tully-Fisher
relation)
range from Milky Way sized ($\sim 200$ km/s) to in one case $\sim 350$ km/s.
However, their absolute magnitudes are somewhat brighter than those
of the Milky Way.    For comparison the Milky Way disk has 
a $B$ band luminosity of $M_B \sim -19.9$  and a disk central
surface brightness of $S_{B} \sim 20.8$ mag/$''^2$
(based on that estimated in $V$ band from \cite{vdk86}; \cite{B+T}, 
and colors typical of an Sc galaxy; \cite{dej96}).

\section{The ``Maximal disk'' Mass-to-Light Ratio}

If the galaxy {\it disk} contributes most of the mass 
in the central few scale lengths 
(meaning that the dark halo contribution is negligible) 
then we can say that the galaxy has a ``maximal disk''.  
Here we estimate the disk mass-to-light ratios of these galaxies 
assuming that they have ``maximal disks''.
We assume here that the disk mass-to-light ratio is constant 
(does not vary with radius).
An exponential disk has a maximum circular velocity at a radius $r=2.2h$ of 
\begin{equation}
v_{max} \approx 0.6 \sqrt{ G M \over h} = 0.6 \sqrt{ G L ~ {M/L} \over h}
\end{equation}
where $L$ is the total luminosity of the disk and $M/L$ is its mass-to-light 
ratio.  If the disk is ``maximal'' then the rotation curve reaches
a maximum near $r=2.2h$ and has approximately 
the circular velocity given by the above equation.
From a prediction or observation of the circular velocity 
we can invert this equation to
determine a ``maximal disk'' mass-to-light ratio given the disk
scale length and absolute magnitude of the galaxy.

For the VFP galaxies we use the observed rotational 
velocities to compute ``maximal disk'' values for the mass-to-light ratio.  
These values are listed in Table 3 along with maximum disk masses 
derived from them.  However, rotational 
velocities have not been observed for the SGGR galaxies.
To compute the maximal disk $M/L$ 
we can use rotational velocities predicted
from the Tully-Fisher relation, however we must do this cautiously
since there is some uncertainty in the offset of the 
Tully-Fisher relation at high redshift.

We estimate errors in the ``maximal disk'' mass-to-light ratio
by considering the uncertainty in the Tully-Fisher relation for distant 
galaxies.  Recent comparisons of the Tully-Fisher relation  between local
and distant galaxies find  a mean offset 
$< 0.4$ mag (Vogt et al. 1997\cite{}) and $1.5$ mag 
(Rix et al. 1997\cite{}) (at $B$ band)
such that distant galaxies of a given magnitude 
have lower velocities than nearby ones ($z=0$).
\cite{vog97} also sees a typical scatter of $\sim 1$ mag for
observed galaxies about the local Tully-Fisher relation.
We compute that an offset of 1 mag in the Tully-Fisher relation results in 
a decrease in our value for $v_{TF}$ by a factor of $0.77$ 
and a resulting maximal disk mass-to-light ratio which is a factor
of $0.58$ times that computed from the local Tully-Fisher relation
(shown in Table 3).
The maximal disk $M/L$ depends on the square of the rotational velocity.
Another comparison can be made by computing the maximal disk $M/L$
for the VFP galaxies by using velocities predicted from the Tully-Fisher
relation (using \cite{pie92}) instead of those observed.   We compute 
that $M/L_B = 3.4, 3.6, 3.3$ for 074-2237, 0305-00115, 064-4442
respectively using $v_{TF}$ derived from the local
Tully-Fisher relation and $q_0 =0.05$. These ratios are somewhat higher
than those computed from the observed velocities.   If the Tully-Fisher
relation for distant galaxies is indeed offset from the local ($z=0$)
relation then the maximal disk mass-to-light ratios computed using
would be lower than than those listed in Table 3 for the SGGR galaxies.
This would make the SGGR mass to light ratios more consistent with the mean
of the VFP galaxies.

We find that for the galaxies considered here ``maximal disk'' 
mass-to-light ratios are low; 
for the SGGR galaxies $M/L_B = 2.5 - 3.6 ~ M_\odot/L_{B\odot}$ (in solar
units) using the local 
Tully-Fisher relation and $M/L_B = 1.4 - 2.1$ using a Tully-Fisher
relation offset by 1 mag.
The ratios for the VFP galaxies are similar with 
$M/L_B =  1.4-3.0$.
These mass-to-light ratios represent maximum values for the
disk $M/L$, since if it is higher than this,
the circular velocity would be higher
than either that observed (for the VFP galaxies) 
or that predicted from the Tully-Fisher relation (for the SGGR galaxies).

Maximal disk mass-to-light ratios for nearby galaxies lie in the range of 
$1 \lta M/L_r \lta 7 ~ M_\odot / L_{r\odot}$
(\cite{sac97}, \cite{beg87}, \cite{bro97}, Kent 1987a,b\cite{}).
Using colors typical of Sc galaxies (\cite{dej96}) this range is
equivalent to $2 \lta M/L_B \lta 15 ~ M_\odot/L_{B\odot}$.
Therefore the distant galaxies studied here have mass-to-light ratios
which lie at the low end of the range of ``maximal disk''
values estimated for the Milky Way and for nearby galaxies.
The ratios of these distant galaxies are low enough as to be 
typical of stellar systems which have undergone a rapid
episode of star formation (e.g. \cite{ken84}).
The maximum disk masses (computed from
the maximal disk $M/L$s and listed in Table 3) are not significantly different 
than that estimated for the Milky Way 
($\sim 6 \times 10^{10} ~ M_\odot$, \cite{sac97})
which has a similar rotational velocity. 


Since our galaxies cover a range of inclinations (see Table 1)
yet have remarkably similar ``maximal disk'' mass-to-light ratios,
it is unlikely that large errors 
are caused by the correction for internal extinction (listed
in Table 2).  

\section{Disk Stability}

The Toomre parameter describing local disk stability is given by 
\begin {equation}
Q \equiv {\kappa \sigma \over 3.36 G \Sigma  }
\end{equation}
(\cite{B+T})
where $\Sigma$ is the disk mass surface density,
$\Sigma = S \times (M/L) $ for $S$ the disk surface brightness, $\kappa$ is
the epicyclic frequency, and $\sigma$ is the stellar velocity dispersion. 
Using the approximation $\kappa \sim \sqrt{2} v_c/r$
for $v_c$ the circular velocity which is exact in the case 
of a flat rotation curve,
and an exponential disk described by a central surface brightness $S_0$ 
and a scale length $h$, ($S(r)= S_0 \exp{-r/h}$) 
we can compute the quantity we denote $Q_a$
(an adjusted $Q$),
\begin {equation}
Q_a \equiv {Q M/L \over \sigma} = 
{\sqrt{2} v_c \over 3.36 G r S_0 \exp{-r/h} }
\end{equation}
where we have grouped the unknowns on the left hand side.
(Here we assume that $v_c$ is either observed or well estimated).

In the images (see Fig.\ 1 and VFP) 
we observe that the spiral structure extends a few scale lengths  for
all galaxies.   We therefore compute $Q_a$ at $r=2.2h$
where the rotation curve is expected to reach a maximum 
and be nearly flat.
This radius is also roughly equivalent to that of the solar
neighborhood with respect to
the Milky Way's scale length (\cite{sac97}), and so we can compare
the solar neighborhood's stellar velocity dispersion to that predicted
using the above inequality given $M/L$ for the galaxies considered here.
In Table 3 we list values computed for $Q_a$
using the observed circular velocities for the VFP galaxies
and the circular velocities predicted from 
the local Tully-Fisher relation for the SGGR galaxies.
We compute $Q_a$ using $M/L_B$ in solar units.
Once again we must consider the possible offset in the Tully-Fisher
relation for distant galaxies.  $Q_a \propto v_c$ so that 
an offset of 1 mag in the Tully-Fisher relation results in 
a decrease in $v_{TF}$ and a resulting decrease in $Q_a$ 
by a factor of $0.77$ for the SGGR galaxies.


Since $Q$ should lie in the range $1 < Q < 2$ we can construct the inequality
\begin{equation}
{1 \over Q_a} < {\sigma \over M/L} < {2 \over Q_a}.
\end{equation}
Using this inequality we ask what value of $M/L$ results
in a range for $\sigma$ similar to that observed in the solar neighborhood?
For the SGGR galaxies and 0074-2237 and 064-4442
with $Q_a(r=2.2h) \sim 0.05 ~ M_\odot/L_{B\odot}$ km$^{-1}$s 
a value of $M/L_B \sim 1.0$, ($0.8$) $M_\odot/L_{B\odot}$ in solar units is required
for the local Tully-Fisher relation (and that offset by 1 mag, respectively)
applied to the SGGR galaxies for $ 20 < \sigma < 40$ km/s 
which would be consistent with the stellar velocity
dispersion in the solar neighborhood.  The radial velocity dispersion
at the solar neighborhood in the galactic plane is 
$\sim 30$ km/s (\cite{wie77}).
For 0305-00115 
which has $Q_a(r=2.2h) \sim 0.027 ~ M_\odot/L_{B\odot}$ km$^{-1}$s 
a somewhat lower $M/L_B=0.5 ~  M_\odot/L_{B\odot}$
is required to yield the same range for the velocity dispersion.

For comparison, using values for the Milky Way from
the range discussed in Sackett (1997)\cite{}, 
$v_c = 200$ km/s, $M_B =-19.9$ mag and $h=2.7$ kpc,
we find $M/L_B$ (maximal disk) $= 4.8 ~ M_\odot /L_{B \odot} $ and 
$Q_a(r=2.2h) = 0.10 ~ M_\odot/L_{B\odot}$  km$^{-1}$s. 
The galaxies considered here have maximal
disk mass to light ratios and $Q_a$ substantially lower
than that of the Milky Way.
A lower $M/L$ and lower dispersion than the Milky Way such 
as might be expected from a young recently formed disk would be consistent
with the above inequalities.  

If the mass-to-light ratio is that of the maximal disk
values $\sim 3 ~ M_\odot/L_{B\odot}$ then the stellar
velocity dispersions would be quite high 
$60 < \sigma < 120$ km/s, ($45 < \sigma < 90$ km/s for 
the offset Tully-Fisher relation) 
for the SGGR galaxies, 0074-2237 and 064-4442. 
For 0305-00115, a maximal disk $M/L$ also yields 
a similar range for the velocity dispersion.
If these disks have a vertical to radial velocity
dispersion ratio similar to that of the Milky Way 
(${\sigma_z \over \sigma_r} \sim 0.5$ \cite{MB}), then they must be nearly
unstable ($Q\sim 1$) so that their vertical velocity dispersions are not 
so high that they would be too thick to show 
fine features such as narrow spiral arms.
(Hydrostatic equilibrium requires ${ \sigma_z \over vc} \sim  {h_z \over r}$
for $h_z$ the disk thickness.)  
One possibility is that these galaxies are ``sub-maximal'' or fall short
of having ``maximal disks''.    This would suggest that
a substantial amount of dark matter is present in the luminous
regions of the galaxy.
Another possibility is that a fraction of
the disk light originates from a thick disk stellar component that
is not strongly affected by the spiral structure.  In this case 
$Q_a$ would be an estimate for only the thin disk component showing
the spiral structure, and a maximal disk $M/L$ could be allowed 
(with mass contributed from both thin and thick components).
This could also be true for the Milky Way, since  
as emphasized by Sackett (1997)\cite{}, the Milky Way could have a 
``maximal disk''.


%

Here we consider the role of gas in the galaxy disk.
We follow \cite{jog84} who considered stability to spiral structure formation 
in a two component (gas and stars) model.
When the gas mass fraction is between a few to 10 percent
then as found by \cite{jog84} the combined stellar and gas
system can be unstable to spiral structure even when the more massive stellar
component alone would be stable.   However this situation only
arises when the stellar component is nearly unstable, or
has a Toomre $Q$ value that is within the inequality we assumed above ($Q<2$).

If on the other hand the gas disk is a significant percentage of the disk mass 
then the stability would be determined by the properties of
the gas disk alone since the gas velocity dispersion is expected to be lower
than the stellar one. 
Then equation (4) depends on the gas mass fraction $f_g \equiv \Sigma_g/\Sigma_*$ 
(where $\Sigma_g$ is the gas surface density and $\Sigma_*$ is the stellar
surface density) and becomes
\begin{equation}
{1 \over Q_a} < {\sigma_g\over M/L ~ f_g} < {2 \over Q_a}
\end{equation}
where $\sigma_g$ is the gas velocity dispersion.
Using this relation, our value of $Q_a = 0.05 ~ M_\odot/L_{B\odot}$ km$^{-1}$s
for most of the galaxies 
and a typical gas velocity dispersion of 10 km/s yields
$ 0.5 > f_g  M/L_B > 0.25 M_\odot/ L_{B\odot}$ 
requiring both low gas densities and rather low
mass-to-light ratios.   
This appears to contradict our assumption (in this
paragraph) that the gas disk is a significant percentage of the disk mass.

\section {Summary and Discussion}

In this paper we concentrate on seven distant 
galaxies with evidence of spiral structure in their HST images.
We compute maximal disk mass-to-light ratios based on observed
rotational velocities for the three VFP galaxies and based
on those predicted from the Tully Fisher for the five galaxies
from SGGR.  Maximal disk mass-to-light ratios 
are low ($M/L_B \sim 1.5-3.5 M_\odot/L_{B \odot}$) but within the range
found for nearby galaxies 
($2 \lta M/L_B \lta 15 M_\odot / L_{B \odot}$), taking into account
uncertainties in the offset of the Tully-Fisher relation 
for the SGGR galaxies without measured rotational velocities.
The maximal disk mass-to-light ratios
are so low as to suggest that the galaxies contain a young, 
rapidly formed stellar population.  
The disk masses computed from the maximal disk $M/L$s are 
similar to that of the Milky Way.

Based on the observed spiral structure we compute a quantity $Q_a$
derived from the Toomre stability parameter $Q$ which
depends on the ratio of $M/L$ to the stellar
velocity dispersion.   $Q_a$ can be constrained
since a disk showing spiral structure should have $1 < Q < 2$.
For the galaxies considered here the ratio of $M/L$ to the velocity dispersion
is lower than that observed in the Milky Way.
For a value of $M/L_B \sim 1$ the velocity dispersions
are similar to that observed in the solar neighborhood.
However, for the maximal disk $M/L$ the velocity dispersions would
be higher than that of the Milky Way and unless the disk is nearly unstable
($Q \sim 1$), the disk would be so thick that
spiral structure should not present as observed.
This suggests that either a fraction of the light originates from
material (such as a thick disk) not strongly affected by the spiral structure,
or there is a substantial amount of dark matter present in the luminous
regions of the galaxy.  

In this paper we have computed some dynamical quantities for a very few
galaxies.  
There are thousands of distant galaxies 
observed at high spatial resolution with HST.
With spectroscopic follow up projects to measure the redshifts 
and rotation properties of these galaxies, it is possible
to compute dynamical quantities 
for a well defined galaxy sample.  A substantial
fraction of these galaxies also show evidence for spiral structure
so that limits can also be placed on the stellar velocity dispersion
(as introduced here).  
Detection of these galaxies in the near-infrared would shed light
on the ages of these disks.  For example, if these galaxies
are very red then, their disks would most likely contain
a larger fraction of old stars, 
have high mass-to-light ratios, and thick disks.
If high quality spectra are observed of individual galaxies
the properties of their stellar populations such 
as their mass-to-light ratios could be
constrained by population synthesis modeling and compared
to values predicted from dynamical arguments.
Near infrared high angular resolution imaging 
(such as is possible with NICMOS) may
constrain what percentage of light originates from an older thick disk. 

Comparison of well defined samples with 
these galaxies should constrain the process of galaxy evolution.  
If the galaxies studied here are similar to faint blue galaxies 
then they could be more typical of later-type, more recently
formed galaxies which are observed during a time of elevated star formation.
However the majority of galaxies seen in the Canada France Hawaii Survey 
(\cite{ham97}, \cite{lil96}) have
red colors typical of star formation occurring slowly over an extended 
period.   It would be informative to look at the morphology of 
galaxies selected with varying colors.

\acknowledgments

We thank the referees for many comments which have improved this paper.
We thank C. Gronwall for providing us with her code which predicts
rest frame luminosities based on the observed $V-I$ color.
We acknowledge helpful discussions and correspondence with 
P. Sackett, C. Gronwall, R. Kennicutt, H.-W. Rix, C. Liu, R. Green, 
G. Rieke and J. Navarro. 
We also acknowledge support from NSF grant AST-9529190 to M. and G. Rieke
and NASA project no. NAG-53359.
The Medium Deep Survey catalog is based on observations with 
the NASA/ESA Hubble Space
Telescope, obtained at the Space Telescope Science Institute, 
which is operated by the
Association of Universities for Research in Astronomy, Inc., 
under NASA contract
NAS5-26555.  The Medium-Deep Survey is funded by STScI grant GO2684.

\clearpage



\begin{deluxetable}{lccccccccccc}
\footnotesize
\tablecaption{Galaxy Parameters}
\tablehead{
\multicolumn{1}{l}{Galaxy} &
\colhead{$z$}              & 
\colhead{$h$}              & 
\colhead{$m_I$}            &
\colhead{$V-I$}            &
\colhead{$i$}              &
\colhead{$v_{observed}$}  \nl
\colhead{}        & 
\colhead{}        & 
\colhead{$''$}    & 
\colhead{mag}     &
\colhead{mag}     &
\colhead{deg}     &
\colhead{km s$^{-1}$}    \nl 
\multicolumn{1}{l}{(1)} &
\colhead{(2)}        & 
\colhead{(3)}        & 
\colhead{(4)}        & 
\colhead{(5)}        & 
\colhead{(6)}        &
\colhead{(7)}        
} 
\startdata
ua400-8    & 0.431 & 0.56  &$19.89 $&$0.83 $ & 56    &    \nl
uem00-4    & 0.477 & 0.97  &$18.61 $&$1.18 $ & 44    &    \nl    
ugk00-1    & 0.275 & 1.05  &$18.12 $&$0.76 $ & 60    &    \nl 
uko01-25   & 0.593 & 0.62  &$20.29 $&$0.62 $ & 36    &    \nl 
uui00-3    & 0.222 & 0.64  &$18.40 $&$0.81 $ & 31    &    \nl 
           &       &       &        &        &       &    \nl 
%
074-2237   & 0.1535 &1.33 &$17.93$& $0.77$   & 80    & $200^{+25}_{-15}$ \nl
0305-00115 & 0.4761 &0.97 &$19.10$& $1.18$   & 46    & $215^{+40}_{-15}$ \nl
064-4442   & 0.8770 &0.46 &$22.07$& $0.99$   & 60    & $200^{+80}_{-40}$ \nl
\enddata
\tablenotetext{}{
NOTES.--  The top five table entries are from SGGR (see also
Ratnatunga et al. 1997), the bottom three from VFP.
Note that 0305-00115 and uem00-4 are the same galaxy.
Columns:
(1) Galaxy identification number;
(2) Redshift;  
(3) Disk exponential scale length.  For the SGGR galaxies this was
estimated from the half light radius;
(4) $I_{814}$ band observed magnitude (error $\sim 0.1$) after correction for galactic
extinction; 
(5) Observed $V-I$ color (error $\sim 0.15$ mag), here $V$ is the HST
F606W filter (mean $5939.6\AA$ and equivalent width $1500.0\AA$)
and $I$ is the F814W filter (mean $7877.5\AA$ and equivalent width $1459.0\AA$);
(6) Galaxy Inclination from the $I_{814}$ images, 
$\cos i = b/a$ (error $\sim 10^\circ$);
(7) Spectroscopically observed circular velocity from VFP.
}
\end{deluxetable}



\begin{deluxetable}{lccccccccccc}
\footnotesize
\tablecaption{Derived Properties}
\tablehead{
\multicolumn{1}{l}{Galaxy}   &
\multicolumn{2}{c}{$h$}      & 
\colhead{$A^i$}              &
\multicolumn{2}{c}{$M_B$}    & 
\colhead{$S_B(r=0)_0$}         &
\colhead{$(B-V)_0$}        &   
\colhead{$(V-I)_0$}         \nl
\colhead{}                & 
\multicolumn{2}{c}{kpc}   & 
\colhead{mag}             &
\multicolumn{2}{c}{mag}   & 
\colhead{mag/$''^{2}$}    & 
\colhead{mag}             &
\colhead{mag}      \nl
\multicolumn{1}{l}{} &
\colhead{$q_0$=0.05} & 
\colhead{$q_0$=0.50} & 
\multicolumn{1}{l}{} &
\colhead{$q_0$=0.05} & 
\colhead{$q_0$=0.50} & 
\multicolumn{1}{l}{} &
\multicolumn{1}{l}{} &  
\multicolumn{1}{l}{} \nl
\multicolumn{1}{l}{(1)} &
\colhead{(2)}        & 
\colhead{(3)}        & 
\colhead{(4)}        & 
\colhead{(5)}        & 
\colhead{(6)}        & 
\colhead{(7)}        & 
\colhead{(8)}        &
\colhead{(9)}        
} 
\startdata
ua400-8    & 2.76  & 2.50  & 0.48  & -20.93 & -20.74 & 19.83 & 0.41 & 0.86 \nl 
uem00-4    & 5.03  & 4.52  & 0.37  & -22.07 & -21.84 & 20.00 & 0.65 & 1.11 \nl    
ugk00-1    & 3.90  & 3.67  & 0.56  & -21.50 & -21.36 & 20.02 & 0.47 & 0.93 \nl 
uko01-25   & 3.57  & 3.13  & 0.17  & -21.12 & -21.83 & 20.21 & 0.17 & 0.61 \nl 
uui00-3    & 2.04  & 1.94  & 0.13  & -20.22 & -20.11 & 19.90 & 0.54 & 1.02 \nl 
           &       &       &       &        &        &       &      \nl 
%
074-2237   & 3.19  & 3.08  & 1.00  & -20.75 & -20.67 & 20.34 & 0.51 & 0.99 \nl
0305-00115 & 5.02  & 4.51  & 0.37  & -22.07 & -21.84 & 20.00 & 0.65 & 1.11 \nl
064-4442   & 3.11  & 2.57  & 0.53  & -20.77 & -20.35 & 20.26 & 0.17 & 0.61 \nl
\enddata
\tablenotetext{}{
NOTES.--  The top five galaxies are from SGGR, the bottom three
from VFP.  Note that 0305-00115 and uem00-4 are the same galaxy.
Columns:
(1) Galaxy identification number;
(2,3) Exponential scale length in kpc for $q_0 = 0.05$ and 
$q_0 = 0.50$ respectively;
(4) Internal extinction correction in the form of Rubin et al. (1985) 
(error $\sim 0.2$ mag);
(5,6) Rest frame ($z=0$) $B$ band absolute magnitude predicted
using C. Gronwall's code using the observed $V-I$ for $q_0 = 0.05$ and
$q_0 = 0.50$ respectively;
(7) Rest frame $B$ band disk central surface brightness;
(8,9) Rest frame $B-V$ and $V-I$ colors predicted by by C. Gronwall's code.
}
\end{deluxetable}



\begin{deluxetable}{lccccccccccc}
\footnotesize
\tablecaption{Dynamical Properties}
\tablehead{
\multicolumn{1}{l}{Galaxy}                &
\multicolumn{2}{c}{$v_{TF}$}              & 
\multicolumn{2}{c}{$M/L_B$   max}         & 
\multicolumn{2}{c}{$M_{disk}$ max }       & 
\multicolumn{2}{c}{$Q_a$}                  \nl 
\colhead{}                                & 
\multicolumn{2}{c}{km s$^{-1}$}           & 
\multicolumn{2}{c}{$M_\odot/L_{B\odot}$}  & 
\multicolumn{2}{c}{$10^{11} M_\odot$}     & 
\multicolumn{2}{c}{$M_\odot/L_{B\odot}$ km$^{-1}$ s}          \nl
\multicolumn{1}{l}{} &
\colhead{$q_0$=0.05} & 
\colhead{$q_0$=0.50} & 
\colhead{$q_0$=0.05} & 
\colhead{$q_0$=0.50} & 
\colhead{$q_0$=0.05} & 
\colhead{$q_0$=0.50} & 
\colhead{$q_0$=0.05} & 
\colhead{$q_0$=0.50} \nl
\multicolumn{1}{l}{(1)} &
\colhead{(2)}        & 
\colhead{(3)}        & 
\colhead{(4)}        & 
\colhead{(5)}        & 
\colhead{(6)}        &    
\colhead{(7)}        &    
\colhead{(8)}        &    
\colhead{(9)}        
} 
\startdata
ua400-8    & 242   &  228   &2.8&2.7 &  1.0    &  0.8    & 0.048  & 0.049  \nl 
uem00-4    & 343   &  320   &3.6&3.4 &  3.8    &  3.0    & 0.043  & 0.045  \nl  
ugk00-1    & 288   &  276   &3.3&3.2 &  2.1    &  1.8    & 0.048  & 0.049  \nl 
uko01-25   & 256   &  234   &3.4&3.2 &  1.5    &  1.1    & 0.055  & 0.058  \nl 
uui00-3    & 194   &  188   &2.5&2.5 &  0.5    &  0.4    & 0.055  & 0.056  \nl 
           &       &        &   &    &         &         &        &        \nl 
%
074-2237   & 229   & 223   &2.6&2.7  &  0.8    &  0.8    & 0.054  & 0.056  \nl 
0305-00115 & 343   & 320   &1.4&1.5  &  1.5    &  1.3    & 0.027  & 0.030  \nl 
064-4442   & 230   & 202   &2.5&3.0  &  0.8    &  0.7    & 0.052  & 0.063  \nl 
\enddata
\tablenotetext{}{
NOTES.--  The top five galaxies are from SGGR, the bottom three
from VFP.  Note that 0305-00115 and uem00-4 are the same galaxy.
Columns:
(1) Galaxy identification number;
(2,3)  Circular velocity predicted using the $B$ band Tully-Fisher relation 
(Pierce \& Tully 1992) and the absolute magnitudes listed in Table 2, 
for $q_0=0.05$ and $q_0 =0.50$ respectively;
(4,5)  $B$ band mass-to-light ratio for a maximal disk computed using Eq.\ 1,  
for $q_0=0.05$ and $q_0 =0.50$ respectively.
Disk parameters used are listed in Table 2;
(6,7)  Maximum disk masses using $M_B$ listed in Table 2 and
$M/L_B$ for a maximal disk,
for $q_0=0.05$ and $q_0 =0.50$ respectively;
(8,9)  $Q_a \equiv Q (M/L_B) /\sigma$ computed using disk parameters 
listed in Table 2, at a radius of $r=2.2h$ (see Eq.\ 3),
for $q_0=0.05$ and $q_0 =0.50$ respectively.
Maximal disk mass-to-light ratios (columns 4,5) and $Q_a$ 
(columns 8,9) were computed
for the five galaxies from SGGR using the circular velocity predicted from
the local Tully-Fisher relation.
For the three VFP galaxies, the spectroscopically measured circular
velocity (see Table 1) was used.
For the SGGR galaxies an offset of 1 mag in the Tully-Fisher 
relation results in
a decrease in the value for $v_{TF}$ and $Q_a$ by a factor of $0.77$
and a maximal disk mass-to-light ration which is a factor
of $0.58$ times that shown here.
}
\end{deluxetable}

\newpage
\begin{figure*}
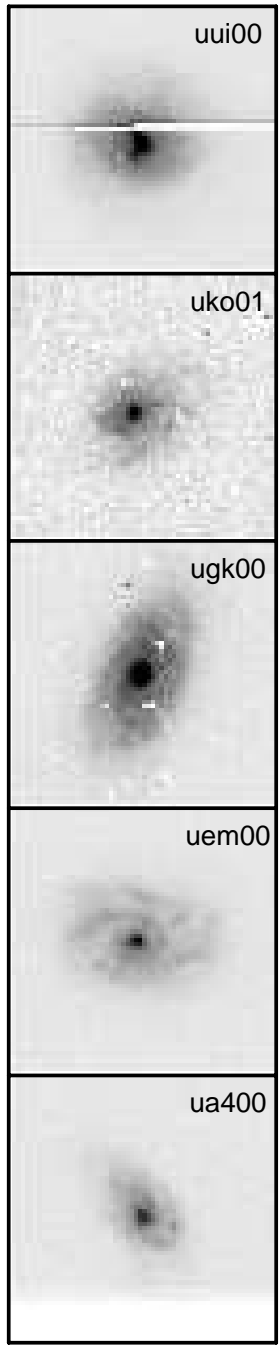

\caption[junk]{
$I_{814}$ images of galaxies with strong spiral structure from SGGR.
Images are $6.4''$ wide and long at a resolution of $0.1''$/pixel.
The horizontal line on the top image is a row of bad pixels.
\label{fig:fig1} }
\end{figure*}
\vfill
 
\end{document}